\documentclass[12pt]{article}%
\usepackage{amssymb}
\usepackage{amsmath}
\usepackage{amsfonts}
\usepackage{sw20ssur}
\usepackage{graphicx}%
\begin{document}

\title{Time evolution of a non-singular primordial black hole}
\author{Manasse R. Mbonye, Nicholas Battista and Benjamin Farr
\and \ \ \\\ \textit{Department of Physics, }\\\textit{Rochester Institute of Technology, }\\\textit{85 Lomb Drive, Rochester, NY 14623.} }
\maketitle

\begin{abstract}
There is growing notion that black holes may not contain curvature
singularities (and that indeed nature in general may abhor such spacetime
defects). This notion could have implications on our understanding of the
evolution of primordial Black holes (PBHs) and possibly on their contribution
to cosmic energy. In this paper we discuss the evolution of a non-singular
black hole (NSBH) based on a recent model [1]. We begin with a study of the
thermodynamic process the black hole in this model and demonstrate the
existence of a maximum horizon temperature $T_{\max}$. At this point the
specific heat capacity $C$ changes signs to positive and the body loses its
black hole characteristics. With no loss of generality, the model is used to
discuss the time evolution of a primordial black hole (PBH), through the early
radiation era of the universe to present, under the assumption that PBHs are
non-singular. In particular, we track the evolution of two benchmark PBHs,
namely the one radiating up to the end of the cosmic radiation domination era,
and the one stopping to radiate currently, and in each case \ determine some
useful features including the initial mass $m_{f}$ and the corresponding time
of formation $t_{f}$. We find that along the evolutionary history of the
universe the distribution of PBH remnant masses (PBH-RM) PBH-RMs follows a
power law. We believe such a result can be a useful step in a study to
establish current abundance of PBH-MRs.

\end{abstract}

\section{Introduction}

Einstein's theory of General Relativity (GR) suggests that under rather
extreme conditions on their density, matter fields can induce physical (or
curvature) singularities on spacetime. Such singularities may be
future-directed as is traditionally assumed to be the case following
gravitational collapse, modelled by the Schwarzschild black hole [2], or they
may be past-directed as in GR-based cosmological solutions [3]. With regard to
the internal dynamics of the matter fields associated with it, the formation
of a spacetime singularity constitutes a "dead end" (when future directed) and
a "dead beginning" (when past directed). Matter just can't wiggle its way out
of a curvature singularity! Thus a "dead beginning", for example, leads to a
paradox of how it is that the initial singularity predicted by GR could have
given rise to the currently observed cosmic dynamics. Such paradoxes have led
to the view that spacetime singularities in GR may simply be a manifestation
of the theory's breakdown at high energy scales. This view, along with the
need to unify GR with Quantum Field Theory (QFT), have led to a search of a
quantum theory of gravity through new frameworks like string theory [4] and
loop quantum gravity [5], (see also [6] for review). While these new
frameworks are very promising, there is still yet no complete theory of
quantum gravity available. In the meantime, an idea which has steadily gained
popularity suggests that at high enough densities [7] regular matter fields
may give way to exotic fields (possibly through phase transitions) which
violate some of the traditional energy conditions so that, for example
$\rho+3p<0$. Such exotic fields are usually modelled with positive definite
energy density $\rho$ and hence with a net negative pressure $p$. It is this
net negative pressure which, through its inflationary effects on spacetime,
offsets the formation of curvature singularities in various non-singular black
a hole models. Understanding and verification of such phase transitions, if at
all, must await formulation of a quantum theory of gravity and/or observations.\ \ \ \ \ \ 

While fields with exotic characteristics as above have never been observed,
their application in physics has not been restricted to black holes par se.
For example, exotic fields have severally been invoked to explain observations
in cosmology. Such invocations range from an inflaton (sitting on a de
Sitter-like potential) initially introduced to explain the horizon problem
using the inflationary paradigm [8] to the various dark energy fields lately
introduced [9] to explain the currently observed late-time cosmic acceleration
[10]. In fact, the practice of invoking exotic fields in physics does date
back quite a while. Recall for example Einstein introduced a cosmological
constant (see for example [11]) in his field equations in order to keep the
universe in static equilibrium from collapse against its own gravity. Later,
[12] Sakharov considered a superdense fluid with an equation of state
$p=-\rho$, and Gliner [13] suggested that such a fluid could constitute the
final state of gravitational collapse. Bardeen's solution [14] was the first
of non-singular black hole models. Since then several solutions of
non-singular black holes have been put forward. Other models have also focused
on regular stars without horizons (gravastars), first suggested by Mazur and
Mottola [15], as possible end-products of gravitational collapse. A review of
various solutions of non-singular spacetimes can be found in [16].\ \ \ \ \ \ \ \ \ \ \ \ \ \ 

When the end-product of gravitational collapse is a black hole, the spacetime
is expected to radiate according to Hawking's prediction [17]. In the
traditional model of a black hole with a curvature singularity the Hawking
radiation process is described by a temperature $T$ that is a monotonically
decreasing function of the black hole mass $T\left(  m\right)  \sim\frac{1}%
{m}$, leading to an increasingly negative specific heat capacity $C$ and
eventual run-away temperatures. Such a body is persistently out of
thermodynamic equilibrium with its surroundings as can be seen (see [20] and
citations therein) from its microscopic entropy $S_{mic}$ dependence on the
(negative) specific heat capacity in $S_{mic}$ $=S-\frac{1}{2}\ln C+...$, with
$S$ being the uncorrected Berkenstein-Hawking entropy. On the other hand when
the end-product of gravitational collapse is a non-singular black hole, the
spacetime will admit two horizons, the exterior Schwarzschild-like horizon and
an interior de-Sitter-like horizon. As such a spacetime radiates, the two
horizons approach each other until they eventually merge [18]. At the point
this merging happens the temperature has a finite maximum [18,19] that
corresponds to a critical mass value. Thereafter the black hole remnant mass
(BH-RM) cools with a positive specific heat capacity and is therefore not a
black hole since it has no horizon. A review of non-singular black hole
radiation process can be found, for example, in Hayward's work [20].\ \ \ \ \ \ \ \ \ 

The possibility that nature may set a (lower bound) length scale so that black
holes are non-singular is one that could have implications on our
understanding of primordial black holes' (PBHs') evolution and their potential
impact on the energy of the universe. PBHs were predicted several decades ago
[21, 22] as topological defects that formed in the early universe. There
exists, in literature, an abundance of mechanisms through which PBHs could
have formed. Such mechanisms include, initial density inhomogeneities,
non-linear metric perturbations, blue spectra of density fluctuations,
equation of state softening, supermassive particles and/or scalar field
dominance and evolution of gravitationally bound objects. A review of these
processes can be found in [23] and citations therein. Note that virtually all
these mechanisms for forming PBHs were specially effective in the very early
universe. As it turns out, PBH formation in the early universe is constrained
within a small time window by the effects of inflation, on the one hand, and
on the other by the requirement to not disrupt nucleosynthesis through high
energy particle emission from Hawking radiation [23].\ There are other
constraints on, for example, the PBH density from cosmic rays [24]. Further,
causality constraints require that at the time of formation a PBH is will be
no larger than the contemporary horizon size. Consequently, only small size
PBHs could form.

In the traditional picture of a singularity containing black hole, there could
\ currently be several different kinds of relics as signatures of PBH [23].
Thus while PBHs of masses $m<10^{15}g$ would have evaporated away, they would
have left behind radiation in form of photons, stable and/or unstable
particles and even naked singularities. Those PBHs with initial masses
$m>10^{15}g$ would still be around. It is also possible that some PBHs could
have survived to seed galaxy formation [25]. On the other hand, if black holes
in general, and PBHs in particular, are non-singular then their evolutionary
path could be somewhat different from the foregoing traditional view. As we
verify, during the evaporation of a non-singular primordial black hole
(NSPBH), the temperature $T\left(  m\right)  $ will evolve to a finite maximum
value at some critical non-zero mass. This (remnant) mass (which as we show is
not a black hole) will then cool down to eventually attain thermodynamic
equilibrium with the surrounding universe. In this case it follows that each
and every PBH that ever formed would leave behind a non-singular primordial
black hole remnant mass (NSPBH-RM) as a signature of its previous existence.
This feature which is generic to NSBH models [26] sets different predictions
of the evolutionary end result of PBHs from those due to traditional singular
models. Observe that since PBHs formed at different times and by implication
with possibly a significant spread in their formation mass spectrum, their
radiation life-times will vary so that the history of the universe should be
littered with creation-events of NSPBH-RMs as end-products of the NSPBHs. One
of the aims of this paper is to inquire on the rate of primordial black hole
remnant mass (PBH-RM) creation during the time evolution of the universe, as a
useful step in the quest towards a count of NSPBH-RMs.\ \ \ \ \ \ 

Finally, it is reasonable to wonder whether such remnant masses from NSPBHs
could contribute any appreciable component to the total energy of the
universe.\ The notion that PBHs could contribute to dark matter has been
raised before (see for example [27]). Considerations for such candidates have,
however, largely focused on the various kinds of relics associated with the
traditional singularity containing black hole, mentioned above. In this paper
the same question is also posed to wonder what contribution, if any,
non-singular end products of PBH evaporation could make to dark matter.
Motivated by this interesting (albeit speculative) scenario we inquire into
the characteristics of the radiation process of NSPBHs. Without loss of
generality, we will utilize a particular model of a non-singular black hole
based on the Mbonye-Kazanas (M-K) solution [1]. For completeness the
discussion starts with a study of thermodynamic processes of this solution. We
then apply the results to PBHs to discuss the PBHs' time evolution during the
various eras of the universe, including the radiation era, and the matter
dominated era.\ Planck units are used in the figures.

This paper is arranged as follows. In section 2 an overview of the solution is
given. In section 3 we develop the thermodynamic features of the M-K solution.
In section 4 the time evolution of the associated PBH is discussed both during
the cosmic radiation era and also continuing into the later, (largely empty)
matter dominated, universe. Section 5 concludes the paper.\ \ \ 

\section{Theoretical framework}

In this section we give an overview of the M-K non-singular black hole
solution [1] which is to be used later to discuss the evolutionary process of
a PBH. The model sketched here, which is an exact solution of the Einstein
field equations, describes the spacetime of a body that gravitationally
collapsed to settle into the final state of a non-singular black hole with
spherical symmetry. The line element takes the form, \ \
\begin{equation}
ds^{2}=-e^{\nu\left(  r\right)  }dt^{2}+e^{\lambda\left(  r\right)  }%
dr^{2}+r^{2}\left(  d\theta^{2}+\sin^{2}\theta d\varphi^{2}\right)  .
\tag{2.1}%
\end{equation}
The traditional black hole with a curvature singularity is represented by the
Schwarzschild solution, which is a vacuum solution. On the other hand an
appropriate non-singular solution must cover both the exterior vacuum region
and the interior matter fields region, and must do so in a way that allows for
some mechanism to offset the singularity that such fields would otherwise
produce. This requires that the various interfaces satisfy matching
constraints, namely the Israel conditions [28]. As is known such constraints
can be quite difficult to satisfy when applied in modeling non-singular black
holes and this can lead to an ill-defined spacetime. Therefore the two main
challenges for this model to overcome are (1) how to introduce gravitating
matter inside the black hole horizon, and at the same time (2) avoid the
creation of a matter-generated curvature singularity.

To deal with these challenges, the M-K model imposes both general and specific
conditions on the fields. The general condition is that the energy momentum
tensor is that of an anisotropic fluid $T_{2}^{2}=T_{3}^{3}\neq T_{1}^{1}$ and
$T_{1}^{1}\neq T_{0}^{0}\neq T_{2}^{2}$. The spacetime is thus Petrov Type
[II, (II)], where ( ) implies a degeneracy in the eigenvalues of the\ Weyl
tensor. An important feature of the model is the specific condition imposed on
the radial pressure $p_{r}=T_{1}^{1}$ as a function of the energy density
$\rho=T_{0}^{0}$ through an equation of state $p_{r}\left(  \rho\right)  $
that takes the form
\begin{equation}
p_{r}\left(  \rho\right)  =\left[  \alpha-\left(  \alpha+1\right)  \left(
\frac{\rho}{\rho_{\max}}\right)  ^{2}\right]  \left(  \frac{\rho}{\rho_{\max}%
}\right)  \rho. \tag{2.2}%
\end{equation}
Here $\rho_{\max}$ corresponds to the maximum density at the core center and
$\alpha$ is a parameter, which when constrained to $\alpha=2.\,\allowbreak
213\,5$ ensures that the sound speed is not super-luminal. It can be seen that
the equation of state above has the required features (see also figure 1).
This equation of state is matter-like at low densities and leads to the
expected dust-like characteristics $p=0$ as $\rho\rightarrow0$. At the very
high density end, deep in the core, the equation asymptotes to that of a
de-Sitter fluid, $p_{r}=-\rho$. It is this outward pressure of a
de-Sitter-like fluid that ultimately offsets the formation of a curvature
singularity. The transition from a matter-like equation of state to eventually
that of a purely de-Sitter fluid at the core is smooth (represented by a
well-behaved function) and appears to suggest the existence of either an
intermediary density dependent quintessential field, $p=w\left(  \rho\right)
\rho,\ -1<w<0$, or a two (matter/de-Sitter) fluid system with a varying
density dependent partial pressure contribution. All in all, it is essential
to emphasize that the continuous nature of the pressure function $p_{r}\left(
\rho\left(  r\right)  \right)  $ represented by Eq. 2.2 gives the net (or
average) value of the pressure due to all fields at a given relevant 2-surface
$r$. Thus for example, the continuous curve AB represents matter that is
growing increasingly relativistic, while CD would represent either a varying
mixture of relativistic matter fields and a de Sitter field or alternatively
represent a radial dependent quintessential field. For details, including how
the various interfaces are smoothly matched, and the interior and exterior
solutions, one can refer to the discussion in [1].\ \ \ %

%TCIMACRO{\FRAME{ftbpFU}{2.373in}{2.1318in}{0pt}{\Qcb{{\small Figure 1: The
%equation of state of matter fields in the MK model. The slope }$\frac
%{dp}{d\rho}${\small \ maximizes\ at point B giveing the speed of light }%
%$c${\small \ as the highest sound speed }$c_{s}${\small \ to avoid
%superluminal behavior. The point C gives max pressure at which a phase
%transition \ in the fields begins.}}}{}{Figure1.eps}%
%{\special{ language "Scientific Word";  type "GRAPHIC";
%maintain-aspect-ratio TRUE;  display "USEDEF";  valid_file "F";
%width 2.373in;  height 2.1318in;  depth 0pt;  original-width 6.5388in;
%original-height 5.8695in;  cropleft "0";  croptop "1";  cropright "1";
%cropbottom "0";  filename 'L9W80G00.eps';file-properties "XNPEU";}}}%
%BeginExpansion
\begin{figure}
[ptb]
\begin{center}
\includegraphics[
height=2.1318in,
width=2.373in
]%
{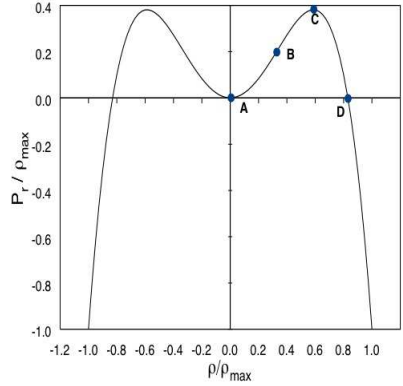}%
\caption{{\small Figure 1: The equation of state of matter fields in the MK
model. The slope }$\frac{dp}{d\rho}${\small \ maximizes\ at point B giveing
the speed of light }$c${\small \ as the highest sound speed }$c_{s}%
${\small \ to avoid superluminal behavior. The point C gives max pressure at
which a phase transition \ in the fields begins.}}%
\end{center}
\end{figure}
%EndExpansion
\ \ \ 

Using the conditions above to solve the Einstein field equations one obtains
an exact solution given by a line element,%

\begin{equation}
ds^{2}=-B(r)dt^{2}+A(r)dr^{2}+r^{2}\left(  d\theta^{2}+\sin^{2}\theta
d\varphi^{2}\right)  , \tag{2.3}%
\end{equation}
where $A\left(  r\right)  =\frac{1}{1-\frac{2GM}{r}}$. For $r<R$, ($R$ being
the size of the total mass $M$) we have (see also [29]) that $B(r)=\exp
\left\{  -\int_{r}^{\infty}\frac{2G}{r^{2}}[m\left(  r^{\prime}\right)  +4\pi
r^{^{\prime}~3}p\left(  r^{\prime}\right)  ]\left[  \frac{1}{1-\frac
{2Gm\left(  r^{\prime}\right)  }{r^{\prime}}}\right]  \right\}  dr^{\prime}$.
For $r\geq R$ the solution takes the familiar form with $B(r)=\frac
{1}{A\left(  r\right)  }=1-\frac{2GM}{r}$.

The tangential pressure $p_{\perp}=\left(  T_{2}^{2}=T_{3}^{3}\right)  $ is
given by\ \
\begin{equation}
p_{\perp}=p_{r}+\frac{r}{2}p_{r}^{\prime}+\frac{1}{2}\left(  p_{r}%
+\rho\right)  \left[  \frac{Gm\left(  r\right)  +4\pi Gr^{3}p_{r}%
}{r-2Gm\left(  r\right)  }\right]  , \tag{2.4}%
\end{equation}
and is a generalization of the Tolman-Oppenheimer-Volkoff equation [30]. Note
that in this model the last term in Eq. 2.4 does not vanish, in general,
except at the center where $p_{r}=-\rho$. This pressure is a unique feature of
the model which in our view is also important. The feature allows the radial
equation of state (Eq. 2.2) to be that of a multi-fluid system, consistently
representing matter fields in the outskirts and a de-sitter field in the
interior core. Note also (from Eqs. 2.2 and 2.4) that at the very center of
the black hole $p_{\perp}=p_{r}=-\rho_{\max}$, and the spacetime is exactly de Sitter.

The spacetime model described above is general enough to admit any density
function $\rho=\rho\left(  r\right)  $ that is a decreasing function of the
radial coordinate. As a particular solution we have taken [18] a density
function of the form $\rho=\rho_{\max}\exp\left[  -\frac{r^{3}}{r_{g}\left(
r_{0}\right)  ^{2}}\right]  $, where $r_{g}=2M=2\int_{0}^{\infty}\rho_{\max
}\{\exp\left[  -\frac{r^{3}}{r_{g}\left(  r_{0}\right)  ^{2}}\right]
\}r^{2}dr$ and $r_{0}=\sqrt{\frac{3}{8\pi G\rho_{\max}}}=\sqrt{\frac
{3}{\Lambda}}$. Then the mass $m\left(  r\right)  $ enclosed by a 2-sphere at
the radial coordinate $r$ is given by $m\left(  r\right)  =\int_{0}^{r}%
\rho_{\max}\{\exp\left[  -\frac{r^{3}}{r_{g}\left(  r_{0}\right)  ^{2}%
}\right]  \}r^{\prime2}dr^{\prime}=M\left[  1-\exp\left(  -\frac{r^{3}}%
{r_{g}\left(  r_{0}\right)  ^{2}}\right)  \right]  $. The entire mass is
essentially concentrated in a region of size $R\simeq(r_{0}^{2}r_{g})^{1/3}$.
Thus, the precise value of $R$ depends on $r_{0}$\ and hence on the value of
$\rho_{\max}$. In this paper we will usually assume (without proof) the
maximum density $\rho_{\max}$ to be of order of the Planck density
$\rho_{pl\text{ }}$and that $r_{0}\sim l_{pl}$, where $l_{pl}$ is the Planck length.

It is easy to verify [1] that the solution given by the line element in Eq.
2.3 leads to the expected asymptotic solutions. Thus for $r>R$ the metric is
described by the vacuum Schwarzschild solution, \
\begin{equation}
ds^{2}=-\left(  \ 1-\frac{2M}{r}\right)  dt^{2}+\frac{1}{\left(  \ 1-\frac
{2M}{r}\right)  }dr^{2}+r^{2}\left(  d\theta+\sin^{2}\theta d\varphi
^{2}\right)  . \tag{2.5}%
\end{equation}
\ On the other hand towards the core as $r\rightarrow r_{0}$, the spacetime
becomes asymptotically de Sitter so that for $0<r\leq r_{0}$,
\begin{equation}
ds^{2}=-\left(  \ 1-\frac{r^{2}}{r_{0}^{2}}\right)  dt^{2}+\frac{1}{\left(
\ 1-\frac{r^{2}}{r_{0}^{2}}\right)  }dr^{2}+r^{2}\left(  d\theta+\sin
^{2}\theta d\varphi^{2}\right)  . \tag{2.6}%
\end{equation}

Such a de-Sitter-like core produces an outward pressure $p=-\rho_{\max}$ which
in turn intervenes against the formation of a curvature singularity inside the
black hole. The result is a non-singular black hole (NSBH).

\section{Thermodynamic features}

\bigskip The spacetime depicted in the solution (Eq. 2.3) has 2 horizons: an
exterior Schwarzschild horizon $r_{+}$ and an interior de-Sitter-like horizon
$r_{-}$. In the neighborhood of each of the horizons, the line element can be
approximated to%

\begin{equation}
ds^{2}=-\left(  \ 1-\frac{\chi\left(  r\right)  }{r}\right)  dt^{2}+\frac
{1}{\left(  \ 1-\frac{\chi\left(  r\right)  }{r}\right)  }dr^{2}+r^{2}\left(
d\theta+\sin^{2}\theta d\varphi^{2}\right)  , \tag{3.1}%
\end{equation}
where $\chi\left(  r\right)  =2m\left(  r\right)  =r_{g}\left[  1-\exp\left(
-\frac{r^{3}}{r_{g}\left(  r_{0}\right)  ^{2}}\right)  \right]  $. The
horizons are given by $r_{+}=\chi_{r\rightarrow r_{g}}$ and $r_{-}%
=\chi_{r\rightarrow r_{0}}$.

\subsection{Hawking temperature}

The Hawking temperature [17] on each of the two horizons is given by
$T=\hbar\kappa_{\pm}\left(  2\pi kc\right)  ^{-1}$.$\ $Here $\kappa_{+}$ and
$\kappa_{-}$are the surface gravity values on the outer and inner horizons,
respectively. To write down the explicit Hawking temperatures for the
spacetime, one first needs an expression for the surface gravity value(s)
$\kappa$. To this end it is convenient to transform Eq. 3.1 (at the horizons)
to Eddington-Finkelstein coordinates so that
\begin{equation}
ds^{2}=-\left(  \ 1-\frac{\chi\left(  r\right)  }{r}\right)  dv^{2}%
+2dvdr^{2}+r^{2}\left(  d\theta+\sin^{2}\theta d\varphi^{2}\right)  ,
\tag{3.2}%
\end{equation}
where $v$ is the advanced time $v=t+r+2m\ln\left\vert r-2m\right\vert $. The
surface gravity $\kappa$ is then obtainable from the relation $l^{a}\nabla
_{a}l^{b}=\kappa l^{b}$ evaluated on the relevant horizon, where $l^{a}$ is
the time-translation killing vector $l^{a}\partial_{a}=\frac{\partial
}{\partial t}$. Taking $l^{a}=\left[  1,0,0,0\right]  $, $l_{a}=-\left[
1-\frac{\chi\left(  r\right)  }{r},1,0,0\right]  $ one finds that $-\frac
{1}{2}\frac{\partial}{\partial r}\left(  -1+\frac{\chi\left(  r\right)  }%
{r}\right)  =\kappa$, giving
\begin{equation}
\kappa_{\pm}=\frac{1}{2}\left[  \frac{\chi}{r_{\pm}}-\frac{\chi^{\prime}%
}{r_{\pm}}\right]  . \tag{3.3}%
\end{equation}

Eqs. 3.3 applied on $T=\hbar\kappa_{\pm}\left(  2\pi kc\right)  ^{-1}$gives
the Hawking temperature $T_{\pm}=\hbar\kappa_{\pm}\left(  2\pi kc\right)
^{-1}$ on the respective horizons as
\begin{equation}
T_{-}=\hbar\left(  4\pi kr_{0}\right)  ^{-1}\left[  \frac{r_{0}}{2m}\left(
1-e^{-4\left(  \frac{m}{r_{0}}\right)  ^{2}}\right)  -6\frac{m}{r_{0}%
}e^{-4\left(  \frac{m}{r_{0}}\right)  ^{2}}\right]  \tag{3.4}%
\end{equation}
and%

\begin{equation}
T_{+}=\hbar\left(  4\pi kr_{0}\right)  ^{-1}\left[  \frac{2m}{r_{0}}-\left(
3+2\frac{2m}{r_{0}}\right)  e^{-\left(  \frac{r_{0}}{2m}\right)  }\right]
\tag{3.5}%
\end{equation}

The dependence of the (outer/inner) horizon temperature $T(m)$ on the mass in
this model is displayed in Figures 2 and 3, respectively. From an observer's
vantage point only the temperature $T_{+}$ of the external horizon will give
readily observable effects. In Figure 2 the solid line shows the evaporation
curve of the non-singular black hole while the dashed line (added for
comparison) shows the evaporation curve of the traditional
singularity-containing black hole. From Figure 2, one can infer that for a
large part of the mass parameter space (down to $\frac{M}{r_{0}}$ $\gtrsim
1.5$) the temperature follows that of the "traditional" singularity-containing
black hole. In this region the specific heat capacity $C\sim\left(
\frac{\partial T}{\partial m}\right)  ^{-1}$ is (as expected) negative and the
black hole, which is out of equilibrium with its environment, gets hotter as
it sheds energy.%

%TCIMACRO{\FRAME{ftbpFU}{2.8461in}{2.1309in}{0pt}{\Qcb{{\small Hawking
%Temperature }$T_{+}${\small \ on the outer horizon as a function of the mass.
%There is a maximum temperature }$T_{\max}${\small \ corresponding to a mass
%}$M_{cr1}${\small . Below this mass the system has a positive specific heat
%capacity (compare with singular bh, dashed line). }}}{}{Figure2.eps}%
%{\special{ language "Scientific Word";  type "GRAPHIC";
%maintain-aspect-ratio TRUE;  display "USEDEF";  valid_file "F";
%width 2.8461in;  height 2.1309in;  depth 0pt;  original-width 6.4714in;
%original-height 4.8395in;  cropleft "0";  croptop "1";  cropright "1";
%cropbottom "0";  filename 'L9W84C01.eps';file-properties "XNPEU";}}}%
%BeginExpansion
\begin{figure}
[ptb]
\begin{center}
\includegraphics[
height=2.1309in,
width=2.8461in
]%
{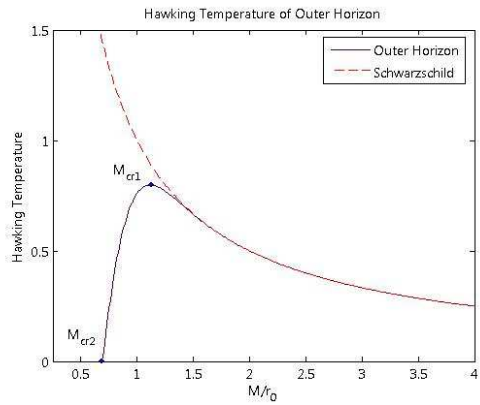}%
\caption{{\small Hawking Temperature }$T_{+}${\small \ on the outer horizon as
a function of the mass. There is a maximum temperature }$T_{\max}%
${\small \ corresponding to a mass }$M_{cr1}${\small . Below this mass the
system has a positive specific heat capacity (compare with singular bh, dashed
line). }}%
\end{center}
\end{figure}
%EndExpansion
%

%TCIMACRO{\FRAME{ftbpFU}{2.8357in}{2.124in}{0pt}{\Qcb{{\small Hawking
%Temperature }$T_{\_}${\small \ on the inner horizon. This is hidden from the
%external observer.}}}{}{Figure3.eps}{\special{ language "Scientific Word";
%type "GRAPHIC";  maintain-aspect-ratio TRUE;  display "USEDEF";
%valid_file "F";  width 2.8357in;  height 2.124in;  depth 0pt;
%original-width 6.4714in;  original-height 4.8395in;  cropleft "0";
%croptop "1";  cropright "1";  cropbottom "0";
%filename 'L9W85802.eps';file-properties "XNPEU";}}}%
%BeginExpansion
\begin{figure}
[ptb]
\begin{center}
\includegraphics[
height=2.124in,
width=2.8357in
]%
{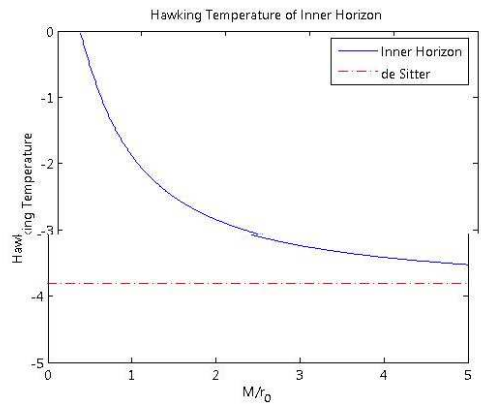}%
\caption{{\small Hawking Temperature }$T_{\_}${\small \ on the inner horizon.
This is hidden from the external observer.}}%
\end{center}
\end{figure}
%EndExpansion
\ \ \ \ 

Eventually the temperature curve begins to depart from that of a black hole
and at some critical mass value $m_{cr1}$ it reaches a maximum $T_{\max}$. In
our model this critical mass value is estimated as $m_{cr1}\approx1.13r_{0}$.
Here the specific heat capacity $C$ vanishes and the microscopic entropy
$S_{mic}$ $=S-\frac{1}{2}\ln C+...$ is no longer well defined. This point
defines a transition in the thermodynamic behavior of the body. Thereafter the
object follows a cooling curve that takes it towards thermodynamic equilibrium
with its environment.

The changes in thermodynamic behavior depicted in Figure 2 are intimately
linked with ongoing changes in gravitational behavior. As the body mass
evaporates towards $m_{cr1}$, the two horizons $r_{+}$ and $r_{-}$ approach
each other. At $m_{cr1}$ the two horizons which have opposite curvatures merge
[18], weakening each others' curvatures in the process and leaving a body with
no horizon. Specifically, while the interior of this remnant mass contains a
de Sitter vacuum, the exterior constitutes matter fields. The repulsive nature
of the interior de Sitter vacuum balances with the attractive nature of the
outside gravitating matter fields to produce a gravitationally stable object.
Such an object is still gravitationally bound, akin to the G-lump first
suggested by Coleman [31] and later discussed by Dymnikova [18]. In fact, to
the extent that the three treatments produce asymptotically de Sitter interior
spacetimes the G-lump and remnant mass should essentially describe the same
object. The only difference then lies in how each of the treatments approaches
this asymptote. In our treatment the pressure of the fields is initially
anisotropic. The stability of the spacetime due to such a G-lump with an
asymptotically de Sitter interior has previously been discussed [32].

One can reasonably speculate that at about the point of transition of the
black hole to the black hole remnant $m_{cr1}$ the matter density $\rho_{cr}$
and temperature $T_{\max}$ are high enough that symmetry restorations of some
of the fundamental forces (at least the electroweak one) has already occurred.
It is therefore natural to wonder whether this, along with the presumption
that the core is de-Sitter-field rich, may observationally obscure the body's
baryonic origins. If so\ could a non-singular black hole remnant mass
(NSBH-RM) be a viable candidate for dark matter?\ 

\subsection{Time evolution}

We now discuss the radiation process of the non-singular black hole in the
model, as set in an empty background. An understanding of this radiation
process is important for two reasons. First, there is the need for
completeness in the evolutionary analysis of the spacetime previously modeled
in [1]. Secondly, the analysis will provide a useful framework in modeling the
time evolution of a PBH in the early radiation dominated universe, represented
by Friedman -Robertson-Walker (FRW) spacetime.

In an empty, asymptotically flat, background the radiation rate of the black
hole can be estimated with use of the Stefan-Boltzman equation,%

\begin{equation}
\frac{dm}{dt}=-\frac{4\pi r_{h}^{2}\sigma T^{4}}{c^{2}}. \tag{3.6}%
\end{equation}
Here $\sigma$ is the Stefan-Boltzman constant. Applying the temperature
function in Eq. 3.5 leads to a mass loss rate of\ \ \ %

\begin{equation}
\frac{dm}{dt}=\frac{-\hbar^{4}G^{2}\sigma m^{2}}{16\pi^{3}k_{B}^{4}c^{10}%
}\left[  \frac{c^{2}}{2Gm}\left(  1-e^{-\left(  \frac{2Gm}{r_{0}c^{2}}\right)
^{2}}\right)  +\frac{6Gm}{r_{0}^{2}c^{2}}e^{-\left(  \frac{2Gm}{r_{0}c^{2}%
}\right)  ^{2}}\right]  ^{4}. \tag{3.7}%
\end{equation}
This equation is separable and, as such, one is tempted to seek analytic,
albeit approximate solutions. However, in order to connect later with the
upcoming analysis of time evolution of PBHs for which the differential
equation will not be separable (during the early radiation era of the
universe), we find it useful to proceed with a numerical approach.

It is convenient to set Eq. 3.7 in a dimensionless form. This is done by
introducing dimensionless quantities $\tilde{m}$ and $\tilde{t}$, respectively
given by $m=m_{c}\tilde{m}=\left(  \frac{r_{0}c^{2}}{2G}\right)  \tilde{m}$
and $t=t_{c}\tilde{t}=\left(  \frac{32\pi^{3}r_{0}^{3}k_{B}^{4}}{\hbar
^{4}G\sigma}\right)  \tilde{t}$. For the purposes of the present calculation,
and as previously mentioned, we will take $r_{0}\sim l_{pl}$. With these
transformations, Eq. 3.7 takes the form
\begin{equation}
\frac{d\tilde{m}}{dt}=-\tilde{m}^{2}\left[  \frac{1}{\tilde{m}}\left(
1-e^{-\tilde{m}^{2}}\right)  -3\tilde{m}e^{-\tilde{m}^{2}}\right]  ^{4}.
\tag{3.8}%
\end{equation}

Figure 4 shows the results of integrating Eq. 3.8. for four different initial
black hole masses.\ It is seen that the time evolution of each of the black
hole masses ends with a non-radiating mass remnant $m_{r}$. This remnant mass,
which is expectedly independent of the initial black hole mass, corresponds to
$m_{cr1}$ in the earlier discussion for $T\left(  m\right)  $. Note that the
time evolution of the two types of black holes is virtually identical, for the
most part. However, at the very end the two models become different when,
unlike in the case for NSBH, the radiation evolution in the
singularity-containing black hole ends in an explosive process with run-away
temperatures.\thinspace

\section{A non-singular primordial black hole (NSPBH)}

It is usually believed that several processes existed [23] in the early
universe that could have led to the formation of primordial black holes (PBHs)
[21,22]. At the time of formation $t_{f}$ a PBH is constrained by causality to
be no larger than the contemporary horizon size of the universe $a_{h}\left(
t_{f}\right)  $. As a result such black holes were generally small and hot. On
the other hand, the reheating period at the end of cosmic inflation is known
to have produced a large amount of radiation which was to dominate the energy
density of the universe for about $3\times10^{5}$ years. This cosmic radiation
domination era (RDE) does affect both the formation and time evolution of PBHs
in the following respects. First, during the RDE the dynamics of the universe
and hence the evolution of the cosmic horizon size $a_{h}$ is determined
(largely) by the available cosmic background radiation energy (CBR) density
$\rho_{rad}$. In turn, as already pointed out, the horizon size will set the
upper limit of the PBH formation mass $m_{f}$ at the formation time $t_{f}$
through the causality constraint. Secondly, because the PBH is bathed in the
CBR it will, from its formation time, accrete this energy.

It follows then that the evolution of a PBH in the early universe is driven by
two competing processes: the loss of mass through Hawking radiation and the
gain of mass through accretion of the CBR. Consequently, a realistic model of
the time evolution of a PBH must take into account the two competing effects.
This implies that the PBH time evolution will generally take the
form,\ \ \ \ \
\begin{equation}
\left(  \frac{dm}{dt}\right)  _{PBH}=\left(  \frac{dm}{dt}\right)
_{Hawking}+\left(  \frac{dm}{dt}\right)  _{accretion} \tag{4.1}%
\end{equation}
Where $\left(  \frac{dm}{dt}\right)  _{Hawking}<0$ and $\left(  \frac{dm}%
{dt}\right)  _{accretion}>0$. Classically, the accretion term is given by
\begin{equation}
\left(  \frac{dm}{dt}\right)  _{accretion}=\sigma_{g}f_{rad}, \tag{4.2}%
\end{equation}
where $\sigma_{g}=\frac{27\pi}{4}r_{g}^{2}$ is the gravitational accretion
cross-section and $f_{rad}=c\rho_{rad}$ is the accreting radiation flux. With
this Eq. 4.1 can be written as \
\begin{equation}
\left(  \frac{dm}{dt}\right)  _{PBH}=f\left(  m\right)  +\frac{27\pi G^{2}%
}{c^{3}}\rho_{rad}m^{2}, \tag{4.3}%
\end{equation}
where $f\left(  m\right)  =\left(  \frac{dm}{dt}\right)  _{Hawking}$ is the
Hawking term in Eq. 3.7.%
%TCIMACRO{\FRAME{ftbpFU}{2.6143in}{2.0842in}{0pt}{\Qcb{{\small Time evolutions
%of a non-singular and sigular BHs compared }}}{}{Figure4.eps}%
%{\special{ language "Scientific Word";  type "GRAPHIC";
%maintain-aspect-ratio TRUE;  display "USEDEF";  valid_file "F";
%width 2.6143in;  height 2.0842in;  depth 0pt;  original-width 6.0848in;
%original-height 4.8447in;  cropleft "0";  croptop "1";  cropright "1";
%cropbottom "0";  filename 'KQSQQX03.eps';file-properties "XNPEU";}}}%
%BeginExpansion
\begin{figure}
[ptb]
\begin{center}
\includegraphics[
height=2.0842in,
width=2.6143in
]%
{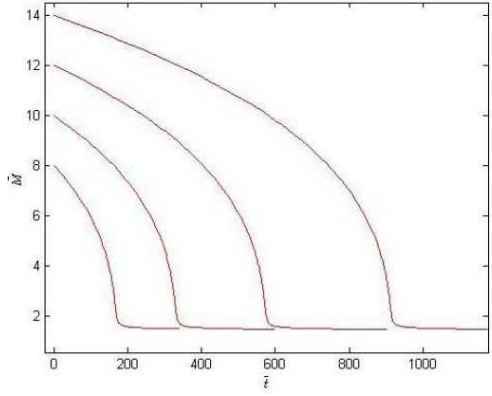}%
\caption{{\small Time evolutions of a non-singular and sigular BHs compared }}%
\end{center}
\end{figure}
%EndExpansion%
%TCIMACRO{\FRAME{ftbpFU}{2.6091in}{1.9969in}{0pt}{\Qcb{{\small Time evolutions
%of a non-singular and sigular BHs compared }}}{}{Figure5.eps}%
%{\special{ language "Scientific Word";  type "GRAPHIC";
%maintain-aspect-ratio TRUE;  display "USEDEF";  valid_file "F";
%width 2.6091in;  height 1.9969in;  depth 0pt;  original-width 6.0995in;
%original-height 4.6613in;  cropleft "0";  croptop "1";  cropright "1";
%cropbottom "0";  filename 'KQSQQX04.eps';file-properties "XNPEU";}}}%
%BeginExpansion
\begin{figure}
[ptbptb]
\begin{center}
\includegraphics[
height=1.9969in,
width=2.6091in
]%
{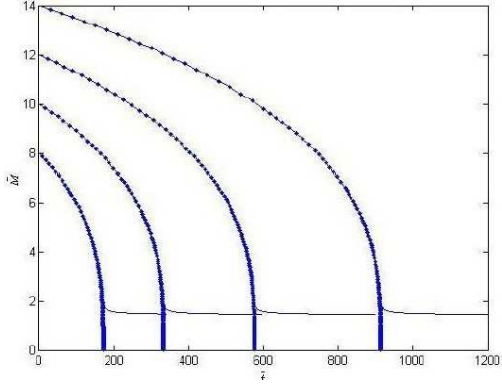}%
\caption{{\small Time evolutions of a non-singular and sigular BHs compared }}%
\end{center}
\end{figure}
%EndExpansion

It is appropriate here to restate the motivating factors for this work. When
one takes the traditional assumption that black holes are singular, one infers
that most of such PBHs ($m<10^{15}g$) would have evaporated away by now, to
leave behind radiation and probably naked singularities. On the other hand, if
black holes are non-singular so that each PBH leaves behind a finite mass
remnant $m_{r}$ at the end of its radiation process, then all PBH remnants
(PBH-RMs) should still be abundant. In the remaining part of this paper we
solve Eq. 4.3 and discuss the results and implications subject to the view
that PBHs are non-singular objects.

\subsection{Background geometry}

Two pieces of information will be needed to solve Eq. 4.3. First, in order to
construct the accretion term $\left(  \frac{dm}{dt}\right)  _{accretion}$
explicitly, one needs the functional dependence of the radiation density on
time, $\rho_{rad}\left(  t\right)  $. Secondly, to perform the integration one
must set up initial conditions in the form of the formation mass $m_{f}$ as a
function of the formation time $t_{f}$. As we show below, these two pieces of
information are non-trivially linked with the black hole background
environment and hence with the background geometry of the contemporary
universe. We briefly review this geometry, which is to be utilized, based on
the Friedman-Robertson-Walker (FRW) model and then utilize it.\ \ 

In the FRW model, the universe is dynamic and its evolution obeys the Einstein
field equations. Under the assumption that the universe is homogeneous and
isotropic, the Einstein equations reduce to the Friedman equations,\ \ \ \
\[
\left(  \frac{\dot{a}}{a}\right)  ^{2}=\frac{8\pi G}{3}\rho-\frac{k}{a^{2}},
\]%
\begin{equation}
\frac{\ddot{a}}{a}=-\frac{4\pi G}{3}\left[  \rho+\frac{3p}{c^{2}}\right]  ,
\tag{4.4}%
\end{equation}
where $a\left(  t\right)  $ is the scale factor, $\rho$ and $p$ are
respectively the density and pressure of the fields, and $k=\left[
1,0,-1\right]  $ is the spacetime curvature parameter. The early universe is
characterized by a nearly flat $\left(  k=0\right)  $ radiation
dominated\ spacetime. On solving the Friedman equations under these conditions
one infers that during the RDE, when $p=\frac{1}{3}c^{2}\rho$, the scale
factor evolves as $a\left(  t\right)  =a_{0}t^{\frac{1}{2}}=t^{\frac{1}{2}}$,
where we take as the current value $a_{0}=1$. It follows then that the
radiation density $\rho_{rad}\sim a^{-4}$ evolves as $\rho_{rad}\sim t^{-2}$.
More explicitly $\rho_{rad}=\frac{\alpha}{\left(  t/s\right)  ^{2}}\ gcm^{-3}%
$, where $\alpha\sim8.4\times10^{4}$.

\subsection{NSPBH\ time evolution}

With $\rho_{rad}\sim\frac{\alpha}{t^{2}}$ Eq. 4.3 can now be written as%
\begin{equation}
\frac{dm}{dt}=f\left(  m\right)  +\frac{27\pi G^{2}\alpha}{c^{3}}\left(
\frac{m}{t}\right)  ^{2}. \tag{4.5}%
\end{equation}
On using the same transformations previously used in Eq. 3.8, a dimensionless
form of Eq. 4.5 takes the form
\begin{equation}
\frac{d\tilde{m}}{dt}=-\tilde{m}^{2}\left[  \frac{1}{\tilde{m}}\left(
1-e^{-\tilde{m}^{2}}\right)  -3\tilde{m}e^{-\tilde{m}^{2}}\right]
^{4}+4.031\times10^{-5}\left(  \frac{\tilde{m}}{t}\right)  ^{2}. \tag{4.6}%
\end{equation}
This equation is clearly not separable and is best solved numerically. To this
end we first set up initial conditions. Recall causality constraints imply
that at its formation time $t_{f}$ a PBH can not be larger than the horizon
size $a_{h}\left(  t\right)  $, defined as $a_{h}\left(  t\right)  =a\left(
t\right)  \underset{t\rightarrow0}{\lim}\int_{t_{i}}^{t}\frac{cdt^{\prime}%
}{a\left(  t\right)  }$. During the radiation era when $a\left(  t\right)
\sim t^{\frac{1}{2}}$, the horizon is given by $a_{h}\left(  t\right)  =2ct$.
The total mass $M_{h}$ of the radiation energy contained in the horizon can
then be estimated as $M_{h}\approx\frac{4}{3}\pi\left(  2ct\right)  ^{3}%
\rho_{rad}=\frac{32}{3}\pi\alpha c^{3}t$. Based on this we will take the upper
bound of the PBH formation mass $m_{f}$ as
\begin{equation}
m_{f}=M_{h}\approx\frac{32}{3}\pi\alpha c^{3}t_{f}. \tag{4.7}%
\end{equation}
Eq. 4.7 provides the initial conditions required to solve Eq. 4.7. Note the
linear relationship between each initial PBH mass and the time of its formation.

As a working example, Figure 6 shows the time evolution of a $1g$\ NSPBH in
the radiation era (solid upper curve). For perspective this is compared with
same initial mass black hole in an empty universe (lower curve). In the model
the growth of the black hole due to the external radiation is initially fast
but not instantaneous (as is seen from a magnification of that part of the
curve). The end of the radiation process is not instantaneous, either (see
also figure 3).

In Figure 7 we display the time evolution of a NSPBH that radiates up to the
end of the universe's radiation era. In Figure 8 we di%
%TCIMACRO{\FRAME{ftbpFU}{2.6238in}{2.2684in}{0pt}{\Qcb{{\small Time evolution
%of 1g NSPBH in radiation era (upper curve) and in empty background (lower
%curve) for comparison.}}}{}{Figure6.eps}%
%{\special{ language "Scientific Word";  type "GRAPHIC";
%maintain-aspect-ratio TRUE;  display "USEDEF";  valid_file "F";
%width 2.6238in;  height 2.2684in;  depth 0pt;  original-width 9.8191in;
%original-height 8.4812in;  cropleft "0";  croptop "1";  cropright "1";
%cropbottom "0";  filename 'L9W8CT05.eps';file-properties "XNPEU";}} }%
%BeginExpansion
\begin{figure}
[ptb]
\begin{center}
\includegraphics[
height=2.2684in,
width=2.6238in
]%
{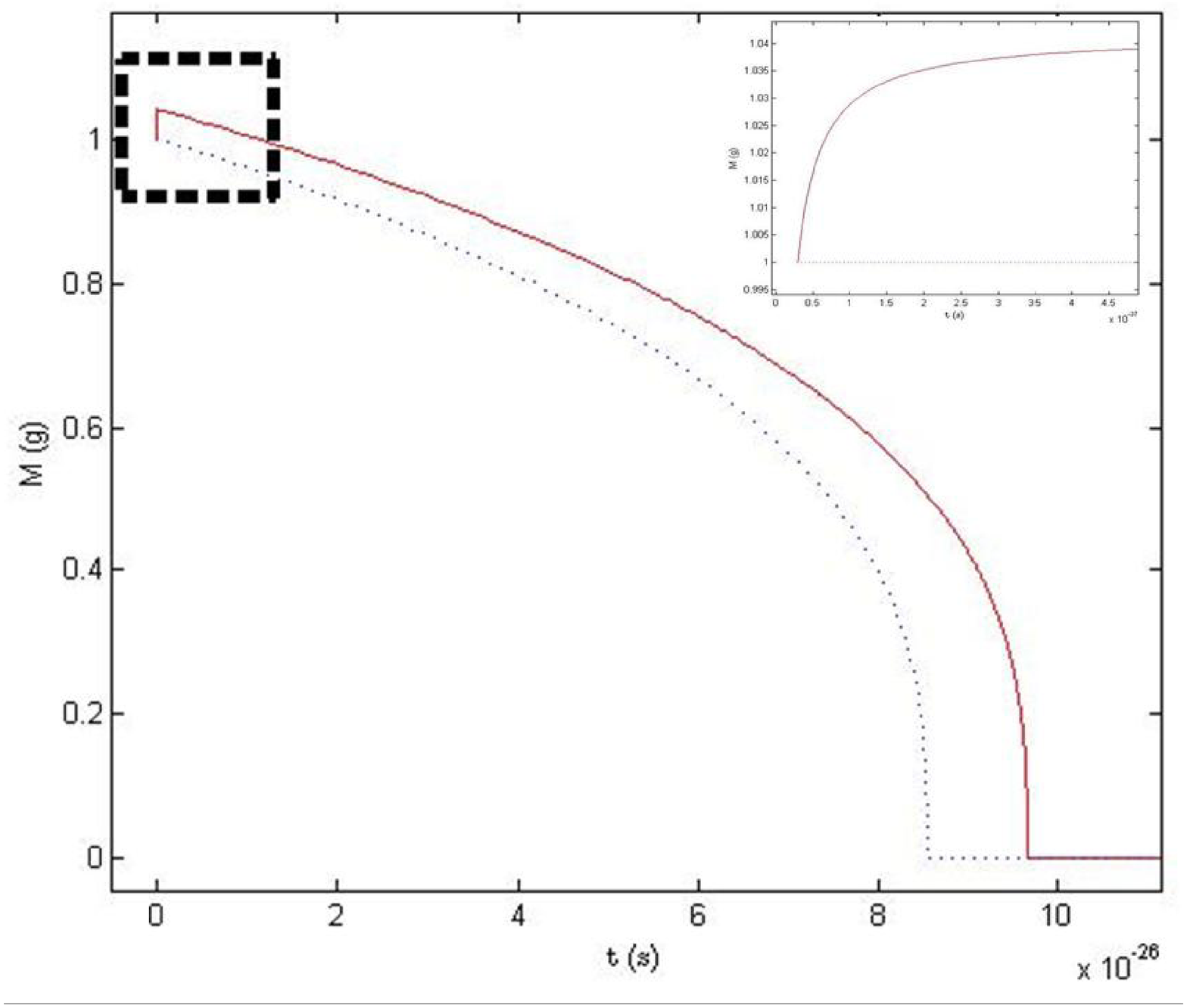}%
\caption{{\small Time evolution of 1g NSPBH in radiation era (upper curve) and
in empty background (lower curve) for comparison.}}%
\end{center}
\end{figure}
%EndExpansion
splay the time evolution of a NSPBH that would be forming a NSPBR-RM now. From
the plots we can infer the following benchmarks : the initial mass $m_{f}%
$\ and time of formation $t_{f}$\ (1) for the PBH that stops radiating to the
end of the radiation era, and (2) for the one that stops radiating now. These
results are tabulated below.

\ 

$%
\begin{array}
[c]{ccc}%
Initial\ mass\ m_{f}\ (gm) & Time\ of\ form\ t_{f}/s &
PBH-RM\ form\ time\ t_{r}/s\\
5.07\times10^{12} & 1.52\times10^{-25} & 1.26\times10^{13}\\
1.65\times10^{14} & 4.95\times10^{-24} & 4.32\times10^{17}%
\end{array}
$

\ 

Further, we find that the total radiation energy $\delta m$\ accreted onto the
PBH during the radiation era is not a significant fraction of the initial mass
at formation $m_{f}$, being of the order of $\delta m\approx0.04m_{f}$. This
is consistent with previous results for singular PBH evolution and evaporation
[33].%
%TCIMACRO{\FRAME{ftbpFU}{2.7475in}{2.2157in}{0pt}{\Qcb{{\small Time evolution
%of a NSBH that radiates up to end of Radiation-Dominated era.}}}%
%{}{Figure7.eps}{\special{ language "Scientific Word";  type "GRAPHIC";
%maintain-aspect-ratio TRUE;  display "USEDEF";  valid_file "F";
%width 2.7475in;  height 2.2157in;  depth 0pt;  original-width 6.9427in;
%original-height 5.591in;  cropleft "0";  croptop "1";  cropright "1";
%cropbottom "0";  filename 'L9W8CT06.eps';file-properties "XNPEU";}}}%
%BeginExpansion
\begin{figure}
[ptb]
\begin{center}
\includegraphics[
height=2.2157in,
width=2.7475in
]%
{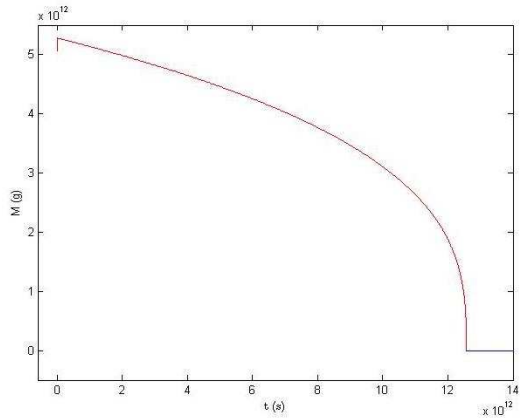}%
\caption{{\small Time evolution of a NSBH that radiates up to end of
Radiation-Dominated era.}}%
\end{center}
\end{figure}
%EndExpansion
$\ $\ \ \ \ \ \ \ 

It is also worthwhile to investigate on the rate at which PBHs die, as a
function of their initial mass $m_{f}$. This information is pa%
%TCIMACRO{\FRAME{ftbpFU}{2.7579in}{2.2018in}{0pt}{\Qcb{{\small Time evolution
%of a NSPBH that ends its radiation process now.}}}{}{Figure8.eps}%
%{\special{ language "Scientific Word";  type "GRAPHIC";
%maintain-aspect-ratio TRUE;  display "USEDEF";  valid_file "F";
%width 2.7579in;  height 2.2018in;  depth 0pt;  original-width 7.0984in;
%original-height 5.6585in;  cropleft "0";  croptop "1";  cropright "1";
%cropbottom "0";  filename 'L9W8CT07.eps';file-properties "XNPEU";}}}%
%BeginExpansion
\begin{figure}
[ptb]
\begin{center}
\includegraphics[
height=2.2018in,
width=2.7579in
]%
{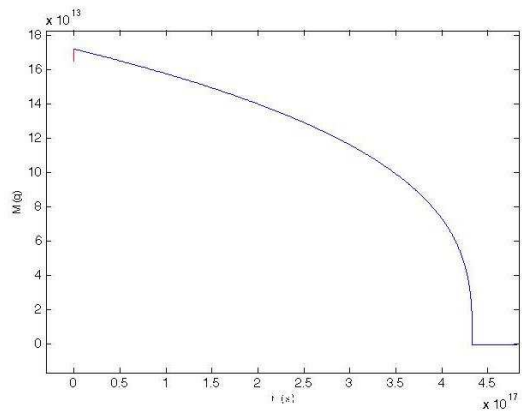}%
\caption{{\small Time evolution of a NSPBH that ends its radiation process
now.}}%
\end{center}
\end{figure}
%EndExpansion
rticularly useful in a non-singular model such as under discussion which
considers the end-product (NSPBH-RM) to be a finite non-black hole mass. One
would like to know how such remnant mass formation events were distributed in
time as the universe evolved. Since each and every PBH that ever existed would
have left behind its remnant mass, it is expected that an integration over
such a mass distribution would lead to a useful description of the current
NSPBH-RM abundances. In turn, modeling NSPBH-RM abundances is a necessary step
in establishing whether NSPBH-RMs could constitute a significant component of
the total energy in the universe. Figure 9 shows the time interval $\Delta
t=t_{f}-t_{r}$ taken for a given PBH to evaporate and form a NSBH-RM as a
function of the initial mass $m_{f}$, for a spectrum of low masses. We find
that this dependence is a power law in the initial mass, $\Delta t\sim
m_{f}^{\gamma}$, where $\gamma\approx3$. \ Further, we have verified (see
Figure 10) that this same power law behavior persists even for larger initial
masses.\ \ \ \ \
%TCIMACRO{\FRAME{ftbpFU}{2.8011in}{2.1119in}{0pt}{\Qcb{{\small Remnant mass
%formation time }$\Delta t_{r}\left(  m_{f}\right)  ${\small \ as a function of
%the formation mass }$m_{f}${\small \ for a spectrum of NSBHs on the large
%}$m_{f}${\small \ end. The power law behavior is same as seen in the low mass
%end.}}}{}{Figure9.eps}{\special{ language "Scientific Word";  type "GRAPHIC";
%maintain-aspect-ratio TRUE;  display "USEDEF";  valid_file "F";
%width 2.8011in;  height 2.1119in;  depth 0pt;  original-width 6.2604in;
%original-height 4.7141in;  cropleft "0";  croptop "1";  cropright "1";
%cropbottom "0";  filename 'L9W8HY08.eps';file-properties "XNPEU";}}}%
%BeginExpansion
\begin{figure}
[ptbptb]
\begin{center}
\includegraphics[
height=2.1119in,
width=2.8011in
]%
{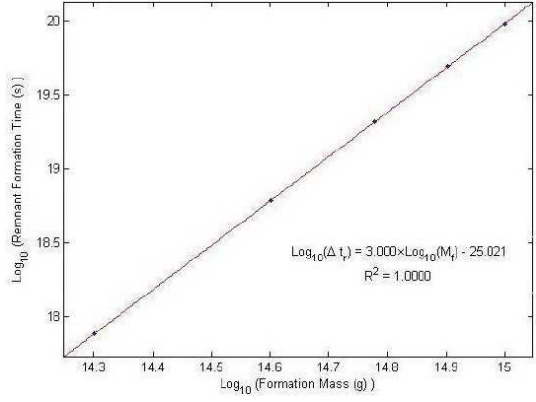}%
\caption{{\small Remnant mass formation time }$\Delta t_{r}\left(
m_{f}\right)  ${\small \ as a function of the formation mass }$m_{f}%
${\small \ for a spectrum of NSBHs on the large }$m_{f}${\small \ end. The
power law behavior is same as seen in the low mass end.}}%
\end{center}
\end{figure}
%EndExpansion

\ \
%TCIMACRO{\FRAME{ftbpFU}{2.8366in}{2.028in}{0pt}{\Qcb{{\small Time interval
%}$\Delta t_{r}=t_{r}-t_{f}${\small \ taken to form a black hole remnant mass
%(NSPBH-RM) as a function of the initial formation mass }$m_{f}${\small \ for a
%spectrum of low masses.}}}{}{Figure10.eps}%
%{\special{ language "Scientific Word";  type "GRAPHIC";
%maintain-aspect-ratio TRUE;  display "USEDEF";  valid_file "F";
%width 2.8366in;  height 2.028in;  depth 0pt;  original-width 6.762in;
%original-height 4.8248in;  cropleft "0";  croptop "1";  cropright "1";
%cropbottom "0";  filename 'L9W8HY09.eps';file-properties "XNPEU";}}}%
%BeginExpansion
\begin{figure}
[ptb]
\begin{center}
\includegraphics[
height=2.028in,
width=2.8366in
]%
{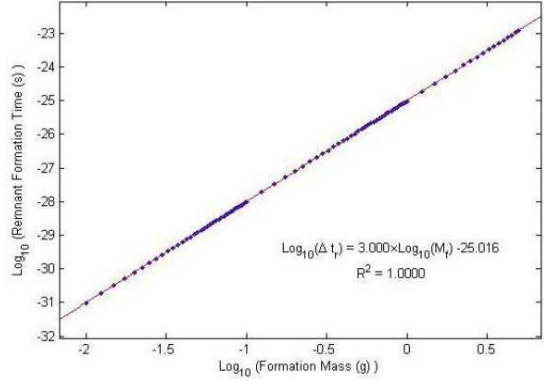}%
\caption{{\small Time interval }$\Delta t_{r}=t_{r}-t_{f}${\small \ taken to
form a black hole remnant mass (NSPBH-RM) as a function of the initial
formation mass }$m_{f}${\small \ for a spectrum of low masses.}}%
\end{center}
\end{figure}
%EndExpansion

{\small Time interval }$\Delta t_{r}=t_{r}-t_{f}${\small \ taken to form a
black hole remnant mass (NSPBH-RM) as a function of the initial formation mass
}$m_{f}${\small \ for a spectrum of low masses.}

\section{Conclusion}

In this paper we have discussed the thermodynamic features and time evolution
of a non-singular black hole (NSBH) based on the Mbonye-Kazanas solution. The
spacetime initially radiates with an increasing temperature driven by a
negative specific heat capacity, reminiscent of a traditional black hole with
a singularity. Eventually, however, the temperature maximizes to $T_{\max}$ at
which point the specific heat capacity $C$ drops to zero and the entropy
$S_{mic}$ $=S-\frac{1}{2}\ln C+...$ of the spacetime is not well defined. It
is here that the spacetime loses its black hole characteristics, including its
horizon. Thereafter the specific heat capacity changes sign becoming positive
and the body cools down as a regular body. We have discussed the time
evolution of this spacetime. Again, here it is shown that the spacetime
radiates to leave a remnant mass (NSBH-RM).

With no loss of generality, we have used the model to investigate the
radiation process of a primordial black hole (PBH), under the assumption that
such a black hole is non-singular. Within this framework, we constructed a
differential equation governing the time evolution of a NSPBH through the
radiation era of the early universe. The equation which is not separable was
integrated numerically to study the time evolution of a PBH. In particular, we
have tracked the evolution of two bench mark PBHs. These include the PBH
radiating up to the end of the radiation domination era and the one stopping
to radiate now. We determined the formation mass $m_{f}$ and the corresponding
formation time $t_{f}$ for each. It is found that the total accreted CBR does
not constitute a significant fraction of the initial PBH mass $m_{f}$ , being
of the order of $0.04m_{f}$ .

Finally, we investigated the rate at which PBHs die, as a function of their
initial mass $m_{f}$. It is found that the rate of primordial black hole
remnant mass (PBH-RM) creation during the time evolution of the universe,
follows a power law of the $\Delta t\sim m_{f}^{\gamma}$, where $\gamma
\approx3$. We believe this result to be a useful spring-point for discussing
current PBH-RM abundances, based on the presumption that black holes are
non-singular. In turn, modeling NSPBH-MR abundances is a necessary step in
establishing whether PBH-remnants could constitute a significant component of
the energy in the universe.\ \ \ \

\end{document}